\RequirePackage{lineno} 

\documentclass[aps,prl,letter,twocolumn,showpacs,superscriptaddress,groupedaddress]{revtex4}  
\usepackage{graphicx}  
\usepackage{dcolumn}   
\usepackage{bm}        
\usepackage{amssymb}   
\usepackage{amsmath}
\usepackage{hyperref}

\newcommand {\Bd} {\ensuremath{B^0_d}}
\newcommand {\Bs} {\ensuremath{B^0_s}}

\newcommand {\Ds} {\ensuremath{D_s^+}}
\newcommand {\Dsm} {\ensuremath{D_s^-}}
\newcommand {\barBd} {\ensuremath{\bar{B}^0_d}}
\newcommand {\barBs} {\ensuremath{\bar{B}^0_s}}

\newcommand {\asld} {\ensuremath{a^d_{\mathrm{sl}}}}
\newcommand {\asls} {\ensuremath{a^s_{\mathrm{sl}}}}

\newcommand {\aslb} {\ensuremath{A^b_{\mathrm{sl}}}}

\newcommand {\ks} {\ensuremath{K^0_S}}

\begin{document}

\hspace{5.2in} \mbox{FERMILAB-PUB-12-338-E}

\title{
Measurement of the semileptonic charge asymmetry using $\bm{\Bs \rightarrow D_{s} \mu X}$ decays
 }
\affiliation{LAFEX, Centro Brasileiro de Pesquisas F\'{i}sicas, Rio de Janeiro, Brazil}
\affiliation{Universidade do Estado do Rio de Janeiro, Rio de Janeiro, Brazil}
\affiliation{Universidade Federal do ABC, Santo Andr\'e, Brazil}
\affiliation{University of Science and Technology of China, Hefei, People's Republic of China}
\affiliation{Universidad de los Andes, Bogot\'a, Colombia}
\affiliation{Charles University, Faculty of Mathematics and Physics, Center for Particle Physics, Prague, Czech Republic}
\affiliation{Czech Technical University in Prague, Prague, Czech Republic}
\affiliation{Center for Particle Physics, Institute of Physics, Academy of Sciences of the Czech Republic, Prague, Czech Republic}
\affiliation{Universidad San Francisco de Quito, Quito, Ecuador}
\affiliation{LPC, Universit\'e Blaise Pascal, CNRS/IN2P3, Clermont, France}
\affiliation{LPSC, Universit\'e Joseph Fourier Grenoble 1, CNRS/IN2P3, Institut National Polytechnique de Grenoble, Grenoble, France}
\affiliation{CPPM, Aix-Marseille Universit\'e, CNRS/IN2P3, Marseille, France}
\affiliation{LAL, Universit\'e Paris-Sud, CNRS/IN2P3, Orsay, France}
\affiliation{LPNHE, Universit\'es Paris VI and VII, CNRS/IN2P3, Paris, France}
\affiliation{CEA, Irfu, SPP, Saclay, France}
\affiliation{IPHC, Universit\'e de Strasbourg, CNRS/IN2P3, Strasbourg, France}
\affiliation{IPNL, Universit\'e Lyon 1, CNRS/IN2P3, Villeurbanne, France and Universit\'e de Lyon, Lyon, France}
\affiliation{III. Physikalisches Institut A, RWTH Aachen University, Aachen, Germany}
\affiliation{Physikalisches Institut, Universit\"at Freiburg, Freiburg, Germany}
\affiliation{II. Physikalisches Institut, Georg-August-Universit\"at G\"ottingen, G\"ottingen, Germany}
\affiliation{Institut f\"ur Physik, Universit\"at Mainz, Mainz, Germany}
\affiliation{Ludwig-Maximilians-Universit\"at M\"unchen, M\"unchen, Germany}
\affiliation{Fachbereich Physik, Bergische Universit\"at Wuppertal, Wuppertal, Germany}
\affiliation{Panjab University, Chandigarh, India}
\affiliation{Delhi University, Delhi, India}
\affiliation{Tata Institute of Fundamental Research, Mumbai, India}
\affiliation{University College Dublin, Dublin, Ireland}
\affiliation{Korea Detector Laboratory, Korea University, Seoul, Korea}
\affiliation{CINVESTAV, Mexico City, Mexico}
\affiliation{Nikhef, Science Park, Amsterdam, the Netherlands}
\affiliation{Radboud University Nijmegen, Nijmegen, the Netherlands}
\affiliation{Joint Institute for Nuclear Research, Dubna, Russia}
\affiliation{Institute for Theoretical and Experimental Physics, Moscow, Russia}
\affiliation{Moscow State University, Moscow, Russia}
\affiliation{Institute for High Energy Physics, Protvino, Russia}
\affiliation{Petersburg Nuclear Physics Institute, St. Petersburg, Russia}
\affiliation{Instituci\'{o} Catalana de Recerca i Estudis Avan\c{c}ats (ICREA) and Institut de F\'{i}sica d'Altes Energies (IFAE), Barcelona, Spain}
\affiliation{Uppsala University, Uppsala, Sweden}
\affiliation{Lancaster University, Lancaster LA1 4YB, United Kingdom}
\affiliation{Imperial College London, London SW7 2AZ, United Kingdom}
\affiliation{The University of Manchester, Manchester M13 9PL, United Kingdom}
\affiliation{University of Arizona, Tucson, Arizona 85721, USA}
\affiliation{University of California Riverside, Riverside, California 92521, USA}
\affiliation{Florida State University, Tallahassee, Florida 32306, USA}
\affiliation{Fermi National Accelerator Laboratory, Batavia, Illinois 60510, USA}
\affiliation{University of Illinois at Chicago, Chicago, Illinois 60607, USA}
\affiliation{Northern Illinois University, DeKalb, Illinois 60115, USA}
\affiliation{Northwestern University, Evanston, Illinois 60208, USA}
\affiliation{Indiana University, Bloomington, Indiana 47405, USA}
\affiliation{Purdue University Calumet, Hammond, Indiana 46323, USA}
\affiliation{University of Notre Dame, Notre Dame, Indiana 46556, USA}
\affiliation{Iowa State University, Ames, Iowa 50011, USA}
\affiliation{University of Kansas, Lawrence, Kansas 66045, USA}
\affiliation{Kansas State University, Manhattan, Kansas 66506, USA}
\affiliation{Louisiana Tech University, Ruston, Louisiana 71272, USA}
\affiliation{Boston University, Boston, Massachusetts 02215, USA}
\affiliation{Northeastern University, Boston, Massachusetts 02115, USA}
\affiliation{University of Michigan, Ann Arbor, Michigan 48109, USA}
\affiliation{Michigan State University, East Lansing, Michigan 48824, USA}
\affiliation{University of Mississippi, University, Mississippi 38677, USA}
\affiliation{University of Nebraska, Lincoln, Nebraska 68588, USA}
\affiliation{Rutgers University, Piscataway, New Jersey 08855, USA}
\affiliation{Princeton University, Princeton, New Jersey 08544, USA}
\affiliation{State University of New York, Buffalo, New York 14260, USA}
\affiliation{University of Rochester, Rochester, New York 14627, USA}
\affiliation{State University of New York, Stony Brook, New York 11794, USA}
\affiliation{Brookhaven National Laboratory, Upton, New York 11973, USA}
\affiliation{Langston University, Langston, Oklahoma 73050, USA}
\affiliation{University of Oklahoma, Norman, Oklahoma 73019, USA}
\affiliation{Oklahoma State University, Stillwater, Oklahoma 74078, USA}
\affiliation{Brown University, Providence, Rhode Island 02912, USA}
\affiliation{University of Texas, Arlington, Texas 76019, USA}
\affiliation{Southern Methodist University, Dallas, Texas 75275, USA}
\affiliation{Rice University, Houston, Texas 77005, USA}
\affiliation{University of Virginia, Charlottesville, Virginia 22901, USA}
\affiliation{University of Washington, Seattle, Washington 98195, USA}
\author{V.M.~Abazov} \affiliation{Joint Institute for Nuclear Research, Dubna, Russia}
\author{B.~Abbott} \affiliation{University of Oklahoma, Norman, Oklahoma 73019, USA}
\author{B.S.~Acharya} \affiliation{Tata Institute of Fundamental Research, Mumbai, India}
\author{M.~Adams} \affiliation{University of Illinois at Chicago, Chicago, Illinois 60607, USA}
\author{T.~Adams} \affiliation{Florida State University, Tallahassee, Florida 32306, USA}
\author{G.D.~Alexeev} \affiliation{Joint Institute for Nuclear Research, Dubna, Russia}
\author{G.~Alkhazov} \affiliation{Petersburg Nuclear Physics Institute, St. Petersburg, Russia}
\author{A.~Alton$^{a}$} \affiliation{University of Michigan, Ann Arbor, Michigan 48109, USA}
\author{G.~Alverson} \affiliation{Northeastern University, Boston, Massachusetts 02115, USA}
\author{A.~Askew} \affiliation{Florida State University, Tallahassee, Florida 32306, USA}
% ...

\collaboration{The D0 Collaboration\footnote{with visitors from
$^{a}$Augustana College, Sioux Falls, SD, USA,
$^{b}$The University of Liverpool, Liverpool, UK,
$^{c}$UPIITA-IPN, Mexico City, Mexico,
$^{d}$DESY, Hamburg, Germany,
$^{e}$SLAC, Menlo Park, CA, USA,
$^{f}$University College London, London, UK,
$^{g}$Centro de Investigacion en Computacion - IPN, Mexico City, Mexico,
$^{h}$ECFM, Universidad Autonoma de Sinaloa, Culiac\'an, Mexico
and
$^{i}$Universidade Estadual Paulista, S\~ao Paulo, Brazil.
}} \noaffiliation
\vskip 0.25cm

\date{July 6, 2012}

\begin{abstract}
We present a  measurement of the time-integrated flavor-specific semileptonic charge asymmetry in  the decays
of \Bs\  mesons that have undergone flavor mixing, \asls , using  $\Bs (\barBs) \rightarrow  D_{s}^\mp \mu^\pm X$ decays, with $D_s^{\mp} \rightarrow  \phi \pi^\mp$ and $\phi \rightarrow K^+ K^-$, using  10.4 fb$^{-1}$ of proton-antiproton collisions collected by the D0 detector during Run II at the Fermilab
Tevatron Collider. 
A fit to the difference between the time-integrated $D_{s}^{-}$ and  $D_s^+$ mass distributions of the \Bs\ and  $\bar{B}^0_s$ candidates 
yields the flavor-specific asymmetry    $ \asls = \left[ \rm{-1.12} \pm  0.74\thinspace (\text{stat}) \pm 0.17 \thinspace (\text{syst}) \right]\%$,
which is the most precise measurement and in agreement with the standard model prediction.
\end{abstract}

\maketitle

CP violation has been observed in the decay and mixing 
of neutral mesons containing strange, charm and bottom quarks. Currently all measurements of CP violation, 
either in decay, mixing or in the interference between the two, have been consistent with the presence of a single phase in the CKM matrix. 
An observation of  anomalously large CP violation in \Bs\ oscillations can indicate the existence of physics beyond the 
standard model (SM)~\cite{smprediction}. 
Measurements of the like-sign dimuon asymmetry by the D0 Collaboration~\cite{dimuon1, dimuon2} show  evidence 
of anomalously large CP-violating 
effects  using data corresponding to 9~fb$^{-1}$ of integrated luminosity.
Assuming that this asymmetry originates from mixed  neutral $B$ mesons, 
the measured value is $\aslb = C_d \asld + C_s \asls = \left[-0.787 \pm 0.172 \, (\text{stat.}) \pm 0.021 \, (\text{syst.}) \right]\%$,
where $a^{s(d)}_{sl}$ is the  time-integrated  
flavor-specific semileptonic charge asymmetry in \Bs (\Bd)  decays  that have undergone flavor mixing 
and $C_{d(s)}$ is the fraction of \Bd (\Bs ) events.  The  value of \asls\ is  extracted from this 
\noindent measurement and found to be
$\asls = (-1.81 \pm 1.06)\%$~\cite{dimuon2}. 
This Letter presents an independent measurement of \asls\ using the  decay 
$\Bs \rightarrow  D_{s}^- \mu^+ X$, where $D_s^{-} \rightarrow  \phi \pi^-$ and $\phi \rightarrow K^+ K^-$ 
(charge conjugate states are assumed in this Letter).

The asymmetry \asls\ is defined as
\begin{equation}
\asls = 
\frac
{\Gamma \left(\barBs \rightarrow \Bs \rightarrow \ell^+ \nu X \right) - \Gamma \left(\Bs \rightarrow \barBs \rightarrow \ell^- \bar{\nu} \bar{X} \right)}
{\Gamma \left(\barBs \rightarrow \Bs \rightarrow \ell^+ \nu X \right) + \Gamma \left(\Bs \rightarrow \barBs \rightarrow \ell^- \bar{\nu} \bar{X} \right)},
\end{equation}
where  in this analysis $\ell=\mu$ and  $X = D_s^{(\ast)-}$. This includes all decay processes of \Bs\ mesons 
that result in a $D_s^-$ meson and an 
oppositely charged muon in the final state.  To study  CP violation, we  identify  events with  the decay
$ \Bs \rightarrow D_s^-  \mu^+ X$. 
The flavor of the \Bs\ meson at the time of decay is identified using the charge of the associated muon, and this analysis does not make use of 
initial-state tagging. The fraction of mixed events integrated over time is extracted using Monte Carlo (MC) simulations.
We assume  there is no production asymmetry between \Bs\ and \barBs\ mesons, 
that there is no direct CP violation in the decay of $D_s$ mesons to the indicated states or in the 
semileptonic decay of $B^0_s$ mesons, and that any CP violation in 
\Bs\ mesons  only occurs in mixing.   We also assume that any direct CP violation in the decay of $b$ baryons and charged $B$ mesons is negligible.
This analysis does not make use of the 
decay $D_s^{-} \rightarrow K^{\ast0}K^-$ as used in Ref.~\cite{d0asls} as the expected statistical 
uncertainty in this channel is 2.5 times worse than the decay  $D_s^- \rightarrow  \phi \pi^-$.

The value of the SM prediction for  
$\asls = \left(1.9 \pm 0.3\right) \times 10^{-5}$~\cite{smprediction}  
is negligible compared with  current experimental precision. The best   direct 
  measurement of \asls\ was performed by the D0  Collaboration using   data corresponding to 5~fb$^{-1}$ 
of integrated luminosity, giving 
$\asls = \left[ -0.17 \pm 0.91 \, (\rm{stat.})
  ^{+0.14}
   _{-0.15}  \, (\rm{syst.})
\right] \%$~\cite{d0asls}.
This Letter presents a new and  improved measurement of \asls\ using the full Tevatron data sample with
an integrated luminosity of  10.4~fb$^{-1}$.

The measurement is performed using the raw asymmetry
 \begin{equation}
 \label{raw}
A = \frac{ N_{\mu^+D_s^{-}} -  N_{\mu^-D_s^{+}}}{ N_{\mu^+D_s^{-}} + N_{\mu^-D_s^{+}}},
\end{equation}
where  $N_{\mu^+D_s^{-}}$ ($N_{\mu^-D_s^{+}}$) is the number of reconstructed $\Bs \rightarrow  \mu^+ D_s^{-} X$ 
($\barBs \rightarrow  \mu^- D_s^{+} X$) decays.  The time-integrated  
flavor-specific semileptonic charge asymmetry in \Bs\ decays which have undergone flavor mixing, \asls ,  is then given by
 \begin{equation}
\asls \cdot {F_{\Bs}^{\text{osc}}} = A - A_{\mu} -A_{\text{track}} - A_{KK} ,
\label{Eq:asls}
\end{equation}
where  $A_{\mu}$ is the reconstruction asymmetry between positive and negatively charged 
muons in the detector~\cite{d0det},  $A_{\text{track}}$
is the   asymmetry between positive and negative tracks, 
$A_{KK}$ is the residual kaon asymmetry from the decay of the $\phi$ meson, and $F_{\Bs}^{\text{osc}}$ is the fraction of  $D_s^- \rightarrow  \phi \pi^-$ 
decays that originate from the decay of a \Bs\ meson after a $\barBs \rightarrow \Bs$ oscillation. 
The $F_{\Bs}^{\text{osc}}$ factor corrects the measured asymmetry for the fraction of events in which the \Bs\ meson is mixed under the assumptions outlined earlier that no other physics asymmetries are present in the other $b$-hadron backgrounds.
While the data selection, fitting models, $A_\mu$,  $A_{\text{track}}$, and $A_{KK}$  were studied, the value of the raw asymmetry was offset by an 
unknown 
arbitrary value and any distribution that  gave an indication of the value of the asymmetry was not examined. 

The D0 detector has a central tracking system, consisting of a 
silicon microstrip tracker (SMT) and a central fiber tracker (CFT), 
both located within a 2~T superconducting solenoidal 
magnet~\cite{d0det, layer0}.  
An outer muon system, at $|\eta|<2$~\cite{eta}, 
consists of a layer of tracking detectors and scintillation trigger 
counters in front of 1.8~T toroidal magnets, followed by two similar layers 
after the toroids~\cite{run2muon}. 

The data are collected with a suite of single and dimuon triggers. 
The selection and  reconstruction of  $\mu^{+} D_s^{-}X$ decays
requires tracks with at least two hits in 
both the CFT and SMT. Muons are required 
to have hits in at least two layers of the muon system,
with   segments reconstructed both  inside and outside the toroid.
The muon track segment has to be matched to a particle found in the central tracking 
system which has momentum $p >  3$~GeV/$c$ and transverse momentum 
$2 < p_T < 25$~GeV/$c$. 

The  $D_s^- \rightarrow  \phi \pi^-$; $\phi \rightarrow K^+ K^-$ decay is reconstructed as follows. 
The two particles from the $\phi$ decay are assumed to be kaons and are required to have  $p_T > 0.7$~GeV/$c$,  
opposite charge and a mass $M(K^+K^-) < 1.07$~GeV/$c^2$. 
The charge of the third particle, assumed to be the charged pion, has to be opposite
       to that of the muon with $0.5 < p_T<25$~GeV/$c$.
The three tracks 
are combined to create a common $D_s^-$ decay vertex using the algorithm described in
Ref.~\cite{vertex}.  To reduce combinatorial background, the $D^-_s$
vertex is required to have a displacement from the $p\bar{p}$ interaction
  vertex  (PV) in
the transverse plane with a significance of at least four standard deviations.  The cosine
of the angle between the $D^-_s$ momentum and the vector from the
PV to the $D^-_s$ decay vertex is required to be greater
than 0.9. The trajectories of the muon and $D^-_s$ candidates are
required to be consistent with originating from a common vertex (assumed to be 
 the $B^0_s$ decay vertex) and to have an effective mass of  $2.6 < M(\mu^+ D_s^-) < 5.4$~GeV$/c^2$, consistent
with coming from a $B^0_s$ semileptonic decay.  The cosine of the angle
between the combined $\mu^+ D^-_s$ direction, an approximation of the
$B^0_s$ direction in the direction from the PV to the
$B^0_s$ decay vertex has to be
greater than 0.95.  The \Bs\ decay vertex has to be displaced from the 
PV in the transverse plane with a significance of at least four standard deviations. 
These angular criteria ensure that the $D^-_s$ and $\mu^+$ momenta are correlated with 
that of their $B^0_s$ parent and that the $D^-_s$ is not  mistakenly  associated with a  random muon. 
 If more than one \Bs\ candidate passes the selection criteria in an event, then all candidates are included in the final sample. 

To improve the significance of the $B^0_s$ selection we use a likelihood ratio
taken  from Refs.~\cite{d0bsmix,like_ratio}. It combines several
discriminating variables: 
the helicity angle between the $D_s^-$ and $K^{+}$ momenta in the  
center-of-mass frame of the $\phi$ meson; 
the isolation of the $\mu^+ D_s^-$ system, defined as $I=p(\mu^+
D_s^-)/[p(\mu^+ D_s^-)+\Sigma p_i]$, where 
$p(\mu^+ D_s^-)$ is the sum of the momenta of the three tracks that make up the $D_s^-$ meson and 
$\Sigma p_i$  is the 
sum of momenta for all   tracks not associated with the $\mu^+ D_s^-$ 
in a cone of $\sqrt{(\Delta \phi)^2+(\Delta \eta)^2}<0.5$ around the
$\mu^+ D_s^-$ direction~\cite{eta};
the $\chi^2$ of the $D_s^-$ vertex fit; 
the invariant masses $M(\mu^+ D_s^-)$, $M(K^+ K^-)$;
and $p_T(K^+ K^-)$.  

The final requirement on the likelihood ratio variable,
$y_{\mathrm{sel}}$, is chosen to maximize the predicted ratio
$N_S/\sqrt{N_S+N_B}$ in a data subsample corresponding to 20\% of the full
data sample, where $N_S$ is the number of signal events and $N_B$ is the
number of background events determined from signal and sideband
regions of the $M(K^+K^-\pi^-)$ distributions.

The $M(K^+K^-\pi^-)$  distribution is analysed
 in bins of 
 6~MeV$/c^2$, 
 over a mass range of $1.7 <  M(K^+K^-\pi^-) < 2.3$~GeV$/c^2$.
The number of events is extracted   by fitting the data to a model using a $\chi^2$ fit.
The $D_s^-$ meson mass distribution is well modelled  by 
two Gaussian functions constrained to have the same  mean, but with 
different  widths and relative normalizations.
A second peak in the $M(K^+K^-\pi^-)$  distribution corresponding to the 
Cabibbo-suppressed decay of the  $D^-$ meson  is  also similarly modelled 
by two Gaussian functions, and the combinatoric 
background  by a third-order  polynomial function. The number of
$D_s^\pm$ signal decays determined from the fit is 
$N(\mu^{\pm}D_s^{\mp}) = 215,\!763 \pm 1,\!467$, where the uncertainty is 
statistical. 

\begin{figure}
\includegraphics[width=\columnwidth]{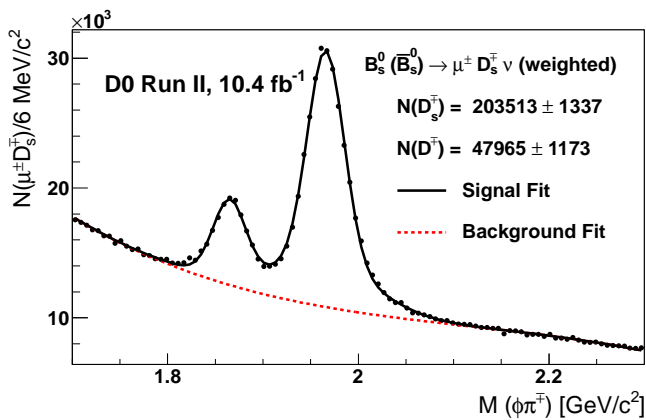}
\caption{\label{Fig:WeightedDSCandidates} 
The weighted $K^+K^-\pi^\mp$ invariant mass distribution for
 the $\mu^{\pm}\phi\pi^\mp$  sample with the  solid line representing the  signal fit and the dashed  line showing the background fit. 
 The lower mass peak is due to  the decay $D^{\mp} \rightarrow \phi \pi^\mp$ and the second peak is due to the $D_s^{\mp}$ meson decay. Note the zero-suppression on the vertical axis.
}
\end{figure}

The polarities of the toroidal and solenoidal magnetic fields are reversed
on average every two weeks so that the four solenoid-toroid polarity
combinations are exposed to approximately the same
integrated luminosity. This allows for a cancellation of first-order
effects related to  instrumental asymmetries.
To ensure full cancellation, the events are weighted according to the number of 
$\mu^+ D_s^-$ decays for each data sample corresponding to a different configuration of the magnets' polarities.
The data are then  fitted  to obtain the number of weighted events,
$N(\mu^\pm D_s^\mp) = 203,\!513 \pm 1,\!337$. This is shown in Fig.~\ref{Fig:WeightedDSCandidates}, where the 
  weighted $M(K^+K^-\pi^-)$ invariant mass distributions in data is compared to the signal and background fit.   

\begin{figure}[htb]
\includegraphics[width=\columnwidth]{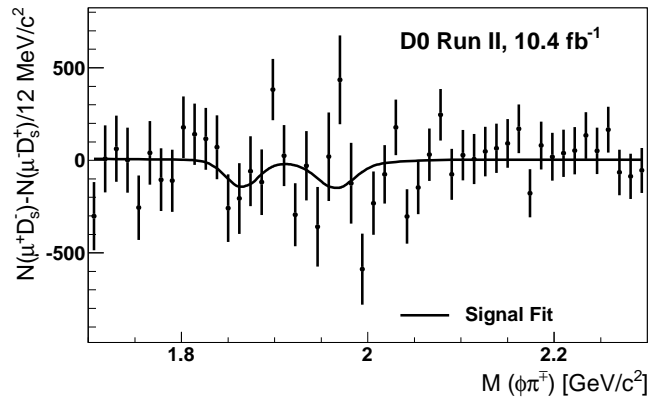}
\caption{\label{Fig:DifferencePlot} 
The fit to the difference distribution for the data (for clarity the data has been rebinned). 
}
\end{figure}

The raw asymmetry  (Eq.~\ref{raw}) is extracted by fitting the $M(\phi\pi^\mp)$ distribution of
the $D_s^\mp$ candidates using a $\chi^2$ minimization. 
The fit is performed simultaneously, using the same models, on the sum (Fig.~\ref{Fig:WeightedDSCandidates}) and
the difference (Fig.~\ref{Fig:DifferencePlot}) of the $M(\phi\pi^-)$ distribution associated with a positively charged muon and $M(\phi\pi^+)$ distribution associated with a negatively charged muon.
The functions $W$ used to model the two distributions are
\begin{eqnarray}
W_{\text{sum}} = & W^{\text{sig}} \left(D_s \right) + W^{\text{sig}} \left(D\right) + W^{\text{bg}}_{\text{sum}}, \\
W_{\text{diff}} = & A W^{\text{sig}} \left(D_s \right) + A_{D}W^{\text{sig}} \left(D\right) +
 A_\text{bg}W^{\text{bg}}_{\text{sum}},
\end{eqnarray}
where $ W^{\text{sig}} \left(D_s \right), W^{\text{sig}} \left(D \right)$ and $W^{\text{bg}}_{\text{sum}}$
describe the $D_s^-$, $D^-$ mass peaks, and the combinatorial background, respectively. 
The asymmetry of the $D^-$ mass peak is $A_{D}$, and $A_\text{bg}$ is the asymmetry of the combinatorial background. 
The result of the fit  is shown in 
Fig.~\ref{Fig:DifferencePlot}  with fitted asymmetry parameters 
$A = (-0.40 \pm 0.33)\%$, $A_{D} = (-1.21 \pm 1.00)\%$, and $A_\text{bg} = (0.00 \pm 0.11)\%$. 

The $\chi^2$ of the fit model with respect to the difference histogram is $129.7/97$ degrees of freedom over the whole mass 
range and    $34.7$ for 25 bins in the mass range $1.90 < M(\mu^+ D_s^{-}) < 2.05$~GeV$/c^2$, which corresponds to a $p$-value of $9.7\%$.
The value of the extracted raw asymmetry, $A$, is checked by calculating the difference between the 
number of $\mu^+\Dsm$ and $\mu^-\Ds$ events in the mass range $1.92 < M(\mu^+ D_s^{-}) < 2.00$~GeV$/c^2$ 
without using a fit. In this region we observe an asymmetry of $(-0.5 \pm 0.3)\%$ which is consistent with the 
value of $A$ extracted by the fitting procedure.

To test the sensitivity of the fitting procedure, the charge of the muon is randomised to introduce an asymmetry signal. 
We use a range of raw signals from $-2.0\%$ to $+2.0\%$ in $0.2\%$ steps with 1000 trials performed for each step,
and the result  of these pseudo-experiments, each with the same statistics as the measurement, is found.
In each case, the central value of the asymmetry 
distribution is consistent with the input value with a fitted width of $0.33\%$ and no observable bias. 
The uncertainty found in  data agrees with this expected statistical sensitivity.

Systematic uncertainties in the fitting method are evaluated by making reasonable variations to the fitting
 procedure.
The mass range of the fit is shifted from
$1.700 <  M(K^+K^-\pi^-) < 2.300$~GeV$/c^2$ to $1.724 <  M(K^+K^-\pi^-) < 2.270$~GeV$/c^2$.
The functions modelling the signal, $W^{\rm sig}$, are modified so that the $D^-$ and $D_s^-$ peaks
  are fitted by single Gaussian functions. 
The background function, $W^{\text{bg}}_{\text{sum}}$, is varied from a 
second-order polynomial function  to a fifth-order polynomial function, and the asymmetry is extracted.
Instead of setting the background of $W_{\text{diff}}$ to $ A_{\text{bg}}W^{\text{bg}}_{\text{sum}}$,
 the background is either set to zero, a constant, or a polynomial function of up to degree three.
The width of the mass bins is varied between 2 and 12~MeV$/c^2$.
 Instead of using the fitted number of \Bs\ decays per magnet polarity to weight the events,  
 the total number of candidates in the 
mass range $1.7 <  M(K^+K^-\pi^-) < 2.3$~GeV/$c^2$ is used.
The systematic uncertainty is assigned to  be 
half of the maximal variation in the asymmetry for each of these sources,   added in quadrature. 
The total effect of all of these systematic sources of uncertainty is a systematic uncertainty of $0.051\%$ on 
the raw asymmetry $A$, giving
\begin{equation}
A = \left[ -0.40 \pm 0.33 \thinspace (\mbox{stat.}) \pm 0.05 \thinspace (\mbox{syst.})\right]\%.
\end{equation}

To extract \asls\ from the raw asymmetry, corrections to the charge asymmetries in the reconstruction have to be made.
These corrections are described in detail in Ref.~\cite{d0adsl}.
The residual detector tracking asymmetry, $A_{\text{track}}$, has been studied in Ref.~\cite{dimuon1} and by using   
 $\ks \rightarrow \pi^+\pi^-$ and
 $K^{\ast\pm} \rightarrow \ks \pi^\pm$ decays. 
 No significant residual track reconstruction
 asymmetries are found and  no  correction for tracking asymmetries need to be  applied.
 The tracking asymmetry of charged pions has been studied using MC simulations of the detector. 
 The asymmetry is found to be less than $0.05\%$, which is assigned as a systematic uncertainty. 
 The muon and the pion have opposite charge, so any remaining  track asymmetries  will cancel   to first order.
 
Any asymmetry between the reconstruction of $K^+$ and $K^-$ mesons  cancels as we require that the two kaons form a $\phi$ meson.
However, there is a small residual asymmetry in the momentum of the kaons produced by the decay of the $\phi$ meson due to
$\phi$-$f_0(980)$ interference~\cite{bellePhi}. The kaon asymmetry is measured using the decay $K^{\ast 0} \rightarrow K^+\pi^-$~\cite{d0adsl}
and is used to determine the residual asymmetry due to this interference, $A_{KK} = \left[ 0.020 \pm 0.002 \thinspace (\mbox{syst.})\right]\%$. 

The  residual reconstruction asymmetry of the muon system, $A_\mu$, has been 
measured using $J/\psi \rightarrow \mu^+\mu^-$ decays 
as described in \cite{dimuon1,dimuon2,d0adsl}. 
This asymmetry is determined as a function of $p_T$ and $|\eta|$ of the muons, 
and the  correction is obtained by a weighted average over the normalized 
yields, as determined from fits to the $M(\phi\pi^-)$ distribution. The resulting correction is  $A_\mu = (0.11 \pm 0.03)\%$ and the combined corrections are 
 $ A_{\mu} + A_{\text{track}} +A_{KK} = \left[ 0.13 \pm 0.06 \thinspace (\mbox{syst.})\right]\%$, 
 including the  statistical uncertainties combined in quadrature.
 
 The remaining variable required is $F_{\Bs}^{\text{osc}}$ (Eq.~\ref{Eq:asls}),  
 which is the only correction extracted from a MC  simulation.  The \Dsm\ 
 signal decays can also be produced via the decay 
of \Bd\, mesons, $B^\pm$ mesons, and from prompt $c\bar{c}$ production. 
 The \Bs\  (\Bd) mesons can oscillate to \barBs\  (\barBd) states before decaying. 
 We split these MC samples into mixed and unmixed decays. 
This classification is inclusive and includes most intermediate excited states
 of both $B$ and $D$ meson decays.
 
 The MC sample is created using the {\sc pythia}  event generator~\cite{pythia} 
 modified to use {\sc evtgen}~\cite{evtgen} for the decay of hadrons containing $b$ and 
 $c$ quarks.  Events recorded in random beam crossings are overlaid over the simulated events 
 to quantify the effect of additional collisions in the same or nearby bunch crossings.
  The {\sc pythia}  inclusive jet production model is used and events are 
 selected that contain at least one muon and a 
 $D_s^- \rightarrow  \phi \pi^-$; $\phi \rightarrow K^+ K^-$ decay. 
 The generated events are processed by the full simulation chain, and then
  by the same reconstruction and selection algorithms as used to select events from  data.
 Each event is  classified
based on the decay chain that is matched to the reconstructed particles.  
 
 The mean proper decay lengths of the $b$-hadrons are fixed in the simulation to values close to
 the current world-average values~\cite{hfag}.
 To correct for these differences, a correction is applied to all non-prompt events in simulation, based on the generated lifetime 
of the $B$ candidate, to give the appropriate world-average $B$ meson lifetimes and measured value of 
the width difference $\Delta\Gamma_s$~\cite{lhcbDeltaGamma}.
 
To estimate the effects of  trigger selection and track reconstruction, we weight each event as 
a function of $p_T$ of the reconstructed muon so that it matches the distribution in the data,  
and as a function of the  lifetime to ensure
 that the $B$-meson lifetimes and $\Delta\Gamma_s$ match the  world-average~\cite{hfag}.

In the case of the
\Bs\ meson, the time-integrated oscillation probability is essentially 50\%  and is insensitive to the  exact 
value of $\Delta M_s$.  
Combining the fraction of  \Bs\ decays in the sample and the time-integrated oscillation 
probability, we find  $F_{\Bs}^{\text{osc}} = 0.465$.

To determine the systematic uncertainty on $F_{\Bs}^{\text{osc}}$, the branching ratios and 
production fractions of $B$ mesons are varied by  their uncertainties.
We also vary the $B$-meson lifetimes and $\Delta\Gamma_s$ and  use a coarser $p_T$ binning in the $p_T$ event weighting.
The total resulting systematic uncertainty on  $F_{\Bs}^{\text{osc}}$  is determined to be $0.017$
that includes the statistical uncertainty from the MC simulation.
An asymmetry of $\Bd$ decays 
of 1\% would contribute  $0.005\%$  to the total asymmetry, which 
 is negligible compared to the statistical uncertainties and 
therefore neglected. 

The uncertainty due to the fitting procedure ($0.05\%$) and the asymmetry corrections 
 ($0.06\%$) are added in quadrature and scaled by the  dilution factor, $F_{\Bs}^{\text{osc}}$. The effect 
of the uncertainty on the dilution factor is then added in quadrature, giving a total systematic 
uncertainty of $0.17\%$.

\begin{figure}[htbp]
\includegraphics[width=\columnwidth]{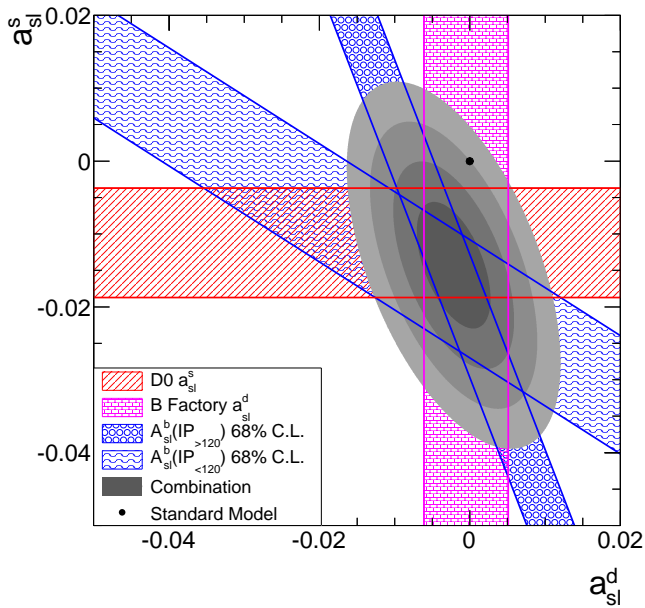}
\caption{\label{Fig:Comparison} 
(color online) A combination of this result 
with two measurements of  \aslb\  with different muon impact parameter selections made using  like-sign dimuons~\cite{dimuon2}
 and the  average  of $\asld$ measurements from $B$ factories~\cite{hfag} . 
The error bands represent the $\pm 1$ standard deviation uncertainties on
each individual measurement.
The ellipses represent the 1, 2, 3, and 4 standard deviation
two-dimensional C.L. regions, respectively, in the 
 $\asls$ and $\asld$ plane.}
\end{figure}

The resulting time-integrated flavor-specific semileptonic charge asymmetry is found to be
\begin{align}
\asls = \left[ \rm{-1.12} \pm 0.74 \thinspace (\text{stat}) \pm 0.17 \thinspace (\text{syst}) \right]\%,
\end{align}
superseding the previous measurement of \asls\ by the D0 Collaboration~\cite{d0asls,comment} and in agreement with the SM prediction.
This result can be combined with the two \aslb\ measurements that depend on the impact parameter of the muons  
(IP)~\cite{dimuon2} and the  average of \asld\ 
measurements from the $B$ factories, $\asld = (-0.05 \pm 0.56)\%$~\cite{hfag},
(Fig.~\ref{Fig:Comparison}).
As a result of this combination we obtain $\asls = (-1.42 \pm 0.57)\%$ and $ \asld = (-0.21 \pm 0.32)\%$
with a correlation of $-0.53$, which is a significant improvement on the
precision  of the measurement of \asld\ and \asls\  obtained in
Ref.~\cite{dimuon2}. These results have
a probability of agreement with the  SM of $0.28\times10^{-2}$, which
corresponds to a 3.0 standard deviations from the SM prediction.

In summary, we have presented the most precise measurement to date of the time-integrated flavor-specific semileptonic charge asymmetry, 
$ \asls = \left[ \rm{-1.12} \pm  0.74\thinspace (\text{stat}) \pm 0.17 \thinspace (\text{syst}) \right]\%$, 
which is in agreement with the standard model prediction and the D0 like-sign dimuon result~\cite{dimuon2}.

%
We thank the staffs at Fermilab and collaborating institutions,
and acknowledge support from the
DOE and NSF (USA);
CEA and CNRS/IN2P3 (France);
MON, NRC KI and RFBR (Russia);
CNPq, FAPERJ, FAPESP and FUNDUNESP (Brazil);
DAE and DST (India);
Colciencias (Colombia);
CONACyT (Mexico);
NRF (Korea);
FOM (The Netherlands);
STFC and the Royal Society (United Kingdom);
MSMT and GACR (Czech Republic);
BMBF and DFG (Germany);
SFI (Ireland);
The Swedish Research Council (Sweden);
and
CAS and CNSF (China).
%


\begin{thebibliography}{99}
 \bibitem{smprediction}
  A. Lenz and U. Nierste,  arXiv:1102.4274;
 A. Lenz and U. Nierste, J. High Energy Phys.  {\bf 06}, 072 (2007). 


\bibitem{dimuon1} 
V.~M. Abazov {\it et al.} (D0 Collaboration), Phys. Rev. D {\bf 82}, 032001 (2010);  
V.~M. Abazov {\it et al.} (D0 Collaboration),   Phys. Rev. Lett. {\bf 105}, 081801 (2010). 

\bibitem{dimuon2} 
V. M. Abazov {\it et al.} (D0 Collaboration), Phys. Rev. D {\bf 84}, 052007 (2011). 
  
 \bibitem{d0asls}
V.~M.~Abazov \emph{et al.} (D0 Collaboration), Phys.\ Rev.\  D {\bf  82}, 012003 (2010).   
 
   \bibitem{d0det}
V.M.~Abazov  \emph{et al.} (D0 Collaboration), Nucl. Instrum. Methods Phys. Res. A {\bf 565}, 463  (2006).

\bibitem{layer0}
R.~Angstadt {\it et al.} (D0 Collaboration),  Nucl. Instrum. Meth. A {\bf 622}, 278  (2010).

\bibitem{eta}
$\eta=-\ln[\tan(\theta/2)]\ $ is the pseudorapidity and $\theta$ is
the polar angle between the track momentum and the proton beam direction. $\phi$ is the azimuthal angle of the track.

 \bibitem{run2muon}  V.M.~Abazov {\it et al.} (D0 Collaboration),  
 	Nucl. Instrum. Meth. A {\bf 552}, 372  (2005).
 
 \bibitem{vertex} 
  J.~Abdallah {\it et al.}  (DELPHI Collaboration),
  Eur.\ Phys.\ J.\ C {\bf 32}, 185 (2004).

 \bibitem{d0bsmix} V.~M.~Abazov {\it et al.} (D0 Collaboration), Phys. Rev. Lett. {\bf 97}, 021802 (2006).

\bibitem{like_ratio}
G.~Borisov,
  Nucl. Instrum. Methods Phys. Res. A {\bf 417}, 384 (1998).

\bibitem {d0adsl} 
V.~M.~Abazov \emph{et al.} (D0 Collaboration),  arXiv:1208.5813, submitted to Phys.\ Rev.\  D.

 
\bibitem{bellePhi}
M.~Stari\v{c} \emph{et al.} (Belle Collaboration), Phys. Rev. Lett. {\bf 108}, 071801 (2012).  
 
 
\bibitem{pythia} 
T.~Sj\"{o}strand, S.~Mrenna and P.~Z.~Skands, J. High Energy Phys. {\bf 05}, 026 (2006).



\bibitem{evtgen} D.G.~Lange, Nucl. Instrum. Methods in Phys. Res. A~{\bf 462}, 152 (2001);
for details see { http://www.slac.stanford.edu/\verb"~"lange/EvtGen}.

  
\bibitem{hfag}
D.~Asner {\it et al.}, Heavy Flavor Averaging Group (HFAG), arXiv:1010.1589;
making use of the 2012 update:  \href{http://www.slac.stanford.edu/xorg/hfag/osc/PDG_2012/}{http://www.slac.stanford.edu/xorg/hfag/osc/PDG\_2012/}

  
\bibitem{lhcbDeltaGamma}
R. Aaij {\it et al.}, (LHCb Collaboration), 	arXiv:1202.4717;
R. Aaij {\it et al.}, (LHCb Collaboration), Phys. Rev. Lett.~{\bf 108}, 101803 (2012). 

\bibitem{comment}
The analysis presented in this Letter has the same statistical uncertainty as the analysis presented in Ref.~\cite{d0asls} 
when performed on the same data sample. 


 
 
\end{thebibliography}
\end{document}